\documentclass[aps,twocolumn,showpacs,superscriptaddress,groupedaddress]{revtex4}  
\usepackage{graphicx}  
\usepackage{dcolumn}   
\usepackage{bm}        
\usepackage{amssymb}   
\usepackage{amsmath}   
\usepackage{feynmp}

\newcommand{\ket}[1]{\left| #1 \right>} 
\newcommand{\bra}[1]{\left< #1 \right|} 

\begin{document}

\widetext
\leftline{Version 2015ww08 as of \today}
\leftline{Comment to {\tt Daniel.Miller@Intel.Com} }


\title{ The Linear Zeeman effect in the molecular positronium Ps2 (dipositronium) }
\author{Daniel~L.~Miller} \affiliation{Intel}

\begin{abstract}
The linear Zeeman effect in the molecular positronium Ps2 (dipositronium) is predicted for
some of $S=1$, $M=\pm1$  states. This result is 
opposite to the case of the positronium atom; the latter has only quadratic 
Zeeman effect. 
\end{abstract}

\pacs{36.10.Dr,32.60.+i,11.30.Er,11.30.Pb}
\maketitle

The lack of the linear Zeeman effect in the positronium is one of the most elementary 
text book problems in QED\cite{IV} reported experimentally as earlier as in 1952\cite{ARich}. 
In the recent years we see return to the positronium research in connection 
with high order radiation correction\cite{alpha7}, antihydrogen synthesis\cite{castelli,aegis}
and positronium molecules\cite{Ps2}. The primary purpose of this work is
to check if the linear Zeeman effect is possible in the dipositronium - the simplest 
neutral system of two electrons and two positrons.

Many authors reported calculations of the stability, symmetry and annihilation of the dipositronium
\cite{Ho,adiabatic,PM,schrader,bailey-frolov,bubin,two-electron} 
using powerful numeric techniques 
and much improving relatively simple calculations of  
Wheeler\cite{wheeler} and Ore\cite{ore}; 
besides the state with $L=1$ was discovered\cite{L-1A,L-1B}. 
The Zeeman effect is not directly calculated in the cited papers. It raises the complexity of the
analysis; besides it was not measured to the best of my knowledge.

The paper shows that the there are two ways to generalize the positronium text book 
solution to the dipositronium; one old (see Sec.~\ref{sec:oldcalc}) 
gives no Zeeman effect similarly to the positronium;
other (new) does predict linear Zeeman effect for four states with $M=\pm1$. 
My argument in favor of the new solution is that it follows 
the decomposition of spinor multiplication.

The four mentioned states of the dipositronium have a non-zero diagonal element
of the spin magnetic moment $\vec \mu$. 
The magnetic field $H$ will shift the energies of these
states by amount
\begin{equation}
    V_H = - \bra{}\mu_z\ket{} H\;,
\end{equation}
that is the linear Zeeman split of quantum energy levels. 
These 4 states have spin $S=1$
and spin projection $M=\pm1$. 
The experiment shows transition between 
$S=0, L=0$ and $S=0, L=1$ energy levels\cite{Ps2}.
Therefore the measurement of the  linear Zeeman effect 
in the mix of the dipositronium and positronium becomes superimportant; 
it will reveal the 
presence of the dipositronium molecules in states with $S=1$.

The spin operator $\vec S$ and the spin magnetic moment $\vec \mu$ 
for the dipositronium molecule are 
\begin{eqnarray}
    \vec S &=& {1\over 2} (\vec \sigma_{e1}  
    + \vec \sigma_{p1}  + \vec \sigma_{e2}  + \vec \sigma_{p2})
    \\
     \vec \mu &=& \mu_0   (\vec \sigma_{p1}  
     - \vec \sigma_{e1}  + \vec \sigma_{p2}  - \vec \sigma_{e2})\;.
\end{eqnarray}
These operators act on the wave function having 
16 components labeled by spins as $\ket{e1,p1,e2,p2}$. 
The eigen-functions of the spin operator $\ket{S,M}$ come in groups
according to eigenvalues of the spin projection $M$. 
The calculation of Clebsch-–Gordan coefficients for system of two positrons and two electrons
should be done in two steps; the first step is the calculation of the decomposition of spinors
\begin{equation}
[1,0]\cdot[0,1]\cdot[1,0]\cdot[0,1] = [0,0]+[2,0]+[0,2]+[2,2]
\end{equation}
the second step is taking the the non-relativistic limit
and transition to the regular spin states.
In other words, the spin states should be classified according to representations of the $SO(4)$.

The 16 combinations of the 4 spins give 9 components to $[2,2]$, 3 components to each 
of $[2,0]$ and $[0,2]$ and last 1 to $[0,0]$. We will label new states by the
spinor index $[J,K]$, by the total spin $S$ and by the spin projection $M$, written
together as $\ket{S,M[J,K]}$.

We start with the Clebsch-–Gordan coefficients for $M=\pm2$; all these states have
zero magnetic moment matrix elements
\begin{eqnarray}
     &&\ket{2,2[2,2]}=\ket{\uparrow\uparrow\uparrow\uparrow}\;,         
      \quad\ket{2,-2[2,2]}=\ket{\downarrow\downarrow\downarrow\downarrow}
 \\     
     && \bra{2,\pm2} \mu_z \ket{2,\pm2} =0\;,
\end{eqnarray}
then continue  with $\ket{S,M=1,[J,K]}$
\begin{equation}
  \begin{array}{c}
      \ket{2,1\text{\small [2,2]}}\\
      \ket{1,1\text{\small [2,2]}}\\
      \ket{1,1\text{\small [1,0]}}\\
     \ket{1,1\text{\small [0,1]}}
   \end{array}
   =
   \left(
  \begin{array}{cccc}
  1/2 & 1/2 & 1/2 & 1/2 \\
  1/2 &- 1/2 & 1/2 &- 1/2 \\
  1/\sqrt{2} & 0 &- 1/\sqrt{2} & 0\\
  0 & 1/\sqrt{2}& 0 & -1/\sqrt{2} \\
   \end{array}
   \right)
       \begin{array}{c}
      \ket{\uparrow\uparrow\uparrow\downarrow}\\
      \ket{\uparrow\uparrow\downarrow\uparrow}\\
      \ket{\uparrow\downarrow\uparrow\uparrow} \\
      \ket{\downarrow\uparrow\uparrow\uparrow} 
   \end{array}
   \label{eq:m=1}
\end{equation}
where numbers in square brackets label the $SO(4)$ original state. Here the breakthrough is coming:
the magnetic moment acquires the diagonal matrix elements
\begin{equation}
  \bra{} \mu_z \ket{} = 2\mu_0
    \left(
  \begin{array}{cccc}
  0 & -1 & 0 & 0 \\
  -1 &0 &0 & 0 \\
  0 & 0 & -1 & 0 \\
  0 & 0 & 0 & 1
   \end{array}
   \right)
\end{equation}
where states should be labeled as in the left hand side of Eq.~(\ref{eq:m=1}). 
Therefore the linear Zeeman split can be observed in $ \ket{1,1\text{\small [1,0]}}$ 
and $ \ket{1,1\text{\small [0,1]}}$ states.
Same result with opposite sign is obtained for $M=-1$; it leads to the linear Zeeman effect for states 
$ \ket{1,-1\text{\small [1,0]}}$ and $ \ket{1,-1\text{\small [0,1]}}$.

The spin states of the dipositronium with $M=0$ are most similar to states of the positronium; the 
magnetic moment vanishes for most of the quantum transitions. To be specific
\begin{eqnarray}
  && \begin{array}{c}
      \ket{1,0\text{\small [0,1]}}\\
      \ket{1,0\text{\small [1,0]}}\\
      \ket{0,0\text{\small [0,0]}}\\
     \ket{2,0\text{\small [2,2]}} \\
     \ket{1,0\text{\small [2,2]}} \\
     \ket{0,0\text{\small [2,2]}} 
   \end{array}
   =  U
          \begin{array}{c}
      \ket{\uparrow\uparrow\downarrow\downarrow}\\
      \ket{\uparrow\downarrow\uparrow\downarrow}\\
      \ket{\uparrow\downarrow\downarrow\uparrow}\\
      \ket{\downarrow\uparrow\uparrow\downarrow}\\
      \ket{\downarrow\uparrow\downarrow\uparrow}\\
      \ket{\downarrow\downarrow\uparrow\uparrow}
 \end{array}
\nonumber   \\
   && U=  {1\over 2}
   \left(
  \begin{array}{cccccc}
  1 & 0& 1 & -1 & 0 & -1 \\
  1 & 0&-1 & -1 & 0 & 1 \\
  1 & 0&-1 & 1 & 0 & -1 \\
  \sqrt{2\over3} &  \sqrt{2\over3}& \sqrt{2\over3} & \sqrt{2\over3} &  \sqrt{2\over3} &  \sqrt{2\over3} \\
  0 &  \sqrt{2} & 0 & 0  & -\sqrt{2} & 0 \\
  {1\over\sqrt{3}}  & -{2\over\sqrt{3}}  & {1\over\sqrt{3}}    &  {1\over\sqrt{3}}  & -{2\over\sqrt{3}}  & {1\over\sqrt{3}}  \\  
   \end{array}
   \right)
   \\
   &&  \bra{} \mu_z \ket{}=  -4\mu_0
   \left(
  \begin{array}{cccccc}
  0 &   &    &  &  &  \\
     & 0&    &  &  &  \\
     &   & 0 &  &  &  \\
     &   &   &  0 & { 1 \over\sqrt{3}} & 0 \\
     &   &   & { 1 \over\sqrt{3}} & 0 &  {\sqrt{2 \over 3}} \\
     &   &   & 0 & {\sqrt{2 \over 3}}   & 0  \\  
   \end{array}
   \right)
\end{eqnarray}
so three $M=0$ states have no Zeemen effect in any order of the perturbation theory; 
three orther $M=0$ states have only the quadratic Zeeman split. 
This similar to the Zeeman effect in positronium\cite{ARich}, where
the dispersion of  $M=\pm1$ states is flat in the magnetic field.

In the summary of the work: the dipositronium magnetic moment is 
calculated for the spin states; the orbital motion of 
the quartet of charges is not discussed. The linear Zeeman effect is found for some of $S=1$ states
and should be accessible experimentally. States with $S=0$ and $S=2$ have no linear shift  in the magnetic field.
These result should help to classify the spin configurations of the dipositronium in real systems.

The calculation of the  Clebsch-–Gordan coefficients for dipositronium in \cite{PM,schrader} was
done in other order; authors take the spin states of positronium and combine them into the dipositronium.
In this scheme the spin magnetic moment has no diagonal elements because it does not have them for 
positronium (see the calculation in the appendix); therefore the linear Zeeman effect does not exists. 
The measurement of the
linear Zeeman split becomes important to understand what is 
the right way to classify the dipositronium spin states.

The dipostronium ground state is assumed to be the $S=0$ singlet by all authors working on this matter. The state
with $S=1$ is therefore the excited state with shorter lifetime and 
harder accessible  for spectroscopy in the magnetic field.
However, the $S>0$ ground state is predicted by the theory 
with anticommuting charge conjugation and particle exchange.\cite{Miller-II}
The ground state should be antisymmetric with respect to permutation 
of the particles and symmetric with respect to permutation
of the antiparticles. For the ground state the spatial wave function should by symmetric, 
so the above symmetry rule is applied to the
spin state. So the system of two particles and two antiparticles must have 
three spins up and one down (or opposite); 
this leads to $M=\pm1$ and  $S=1$ ground state.
The observation of  the linear Zeeman effect in the dipositronium ground state 
will provide data in favor of this theory.

\appendix
\section{The previously published calculation}
\label{sec:oldcalc}

The other ``common'' way to solve this problem
is to think of di-positronium as if it is made from
two positronium atoms having spins $S_1$ and $S_2$.
In this case the states of the dipositronium
are labeled by $\ket{S,M(S_1,S_2)}$, where 
$S$  and $M$ are spin and spin projection of the 
dipositronium.

In this case the Clebsch-–Gordan coefficients
for $M=1$ states (candidates for 
non-zero linear Zeeman effect) are
\begin{equation}
  \begin{array}{c}
      \ket{2,1\text{\small (1,1)}}\\
      \ket{1,1\text{\small (1,1)}}\\
      \ket{1,1\text{\small (1,0)}}\\
     \ket{1,1\text{\small (0,1)}}
   \end{array}
   =
   \left(
  \begin{array}{cccc}
  1/2 & 1/2 & 1/2 & 1/2 \\
  1/2 & 1/2 & -1/2 &- 1/2 \\
  1/\sqrt{2} & - 1/\sqrt{2} \\
  & & 1/\sqrt{2}& -1/\sqrt{2} 
   \end{array}
   \right)
       \begin{array}{c}
      \ket{\uparrow\uparrow\uparrow\downarrow}\\
      \ket{\uparrow\uparrow\downarrow\uparrow}\\
      \ket{\uparrow\downarrow\uparrow\uparrow} \\
      \ket{\downarrow\uparrow\uparrow\uparrow} 
   \end{array}
\end{equation}
and the magnetic moment has no diagonal matrix 
elements similarly to that of the positronium:
\begin{equation}
  \bra{} \mu_z \ket{} = 2\mu_0
    \left(
  \begin{array}{cccc}
  0 & 0 & - 1/\sqrt{2}  & - 1/\sqrt{2} \\
  0 & 0 & - 1/\sqrt{2}  &   1/\sqrt{2} \\
  - 1/\sqrt{2}  & - 1/\sqrt{2} & 0& 0 \\
  - 1/\sqrt{2}  &   1/\sqrt{2} & 0 & 0 
   \end{array}
   \right)
\end{equation}
I believe that this approach is fundamentally wrong, 
because there is no reason for the dipositronium
molecule to preserve the internal spin structure
of each of the positronium atoms.

\end{document}